\documentclass[aps,epsfig,floats,twocolumn,amssymb,amsmath,showpacs]{revtex4}
\usepackage{graphicx}
\begin{document}
\title{Molecular Dynamics Simulation of Semiflexible Polyampholyte Brushes
 - The Effect of Charged Monomers Sequence}

\author{M. Baratlo}

\author{H. Fazli}
\email{fazli@iasbs.ac.ir} \affiliation{Institute for Advanced
Studies in Basic Sciences (IASBS), P. O. Box 45195-1159, Zanjan
45195, Iran}

\date{\today}

\begin{abstract}

Planar brushes formed by end-grafted semiflexible polyampholyte
chains, each chain containing equal number of positively and
negatively charged monomers is studied using molecular dynamics
simulations. Keeping the length of the chains fixed, dependence of
the average brush thickness and equilibrium statistics of the
brush conformations on the grafting density and the salt
concentration are obtained with various sequences of charged
monomers. When similarly charged monomers of the chains are
arranged in longer blocks, the average brush thickness is smaller
and dependence of brush properties on the grafting density and the
salt concentration is stronger. With such long blocks of similarly
charged monomers, the anchored chains bond to each other in the
vicinity of the grafting surface at low grafting densities and
buckle toward the grafting surface at high grafting densities.

\end{abstract}

\maketitle

\section{Introduction}  \label{intro}

Polymers carrying ionizable groups dissolved in a polar solvent
dissociate into charged polymer chains and counter-ions (ions of
opposite charge). Depending on acidic or basic property of
ionizable monomers, charged polymers can be classified into
polyelectrolytes containing a single sign of charged monomers and
polyampholytes bearing both positive and negative charges. These
macromolecules are often water-soluble and have numerous
applications in industry and medicine. Many biological
macromolecules such as DNA, RNA, and proteins are charged
polymers.

Charged monomers of different sign can be distributed randomly
along a polyampholyte chain or charges of one sign can be arranged
in long blocks. With excess of one type of charges, polyampholyte
solutions exhibit properties similar to those of polyelectrolyte
solutions. With the same ratio of positive and negative charges on
the chains, the solution behavior depends noticeably on the charge
sequence. For example, it has been shown that the sequence of
charged amino acids (charge distribution) along ionically
complementary peptides affect the aggregation behavior and
self-assembling process in the solution of such peptides
\cite{Hong,Jun}. Also, using Monte Carlo simulations it has been
shown that charged monomers sequence of neutral
block-polyampholytes affect their adsorption properties to a
charged surface \cite{Messina}.

The present understanding of, in particular, random polyampholytes
in solution and their interaction with surfaces and
polyelectrolytes has recently been reviewed \cite{rev_Dobrynin}.
In addition, theoretical and computer
 simulation studies of single diblock polyampholytes
 \cite{Imbert,Baumketner,Wang},
and diblock polyampholytes in solution
\cite{Castelnovo,Shusharina1}, have been performed.
\begin{figure}[b]
\includegraphics[width=0.69 \columnwidth]{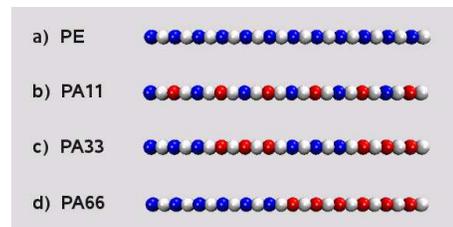}
\caption{(Color online) Schematic structures of four different
chains containing different sequences of charged monomers.
Neutral, positively and negatively charged monomers are shown by
white, red (dark) and blue (black) spheres respectively. The chain
containing only negatively charged monomers is a polyelectrolyte
chain, PE, and three other chains PA11, PA33 and PA66 are
polyampholytes consisting of alternating blocks of similarly
charged monomers with 1, 3 and 6 monomers in each block
respectively.} \label{structures}
\end{figure}
The value of salt concentration in the solution is an important
parameter for tuning the structural properties of polyampholyte
chains. With increasing the salt concentration, the coil size and
the viscosity increase in the solution of nearly charge-balanced
polyampholyte chains and decrease in the case of the solution of
polyampholytes with a large net charge (the polyelectrolyte
regime)\cite{Zheng,Ehrlich,McCormick,Corpart}.

The properties of the system of polymers anchored on a surface are
of great interest both in industrial and biological applications
and academic research. When there is sufficiently strong repulsion
between the polymers, the chains become stretched and the
structure obtained is known as a polymer brush. Brushes at planar
and curved surfaces formed by grafted homopolymers have
extensively been investigated by various theoretical methods
\cite{Wijmans,Lindberg,Klos,Almusallam}.

In the case of polyelectrolytes grafted onto a surface
(polyelectrolyte brush),
 the repulsion of electrostatic origin between the
chains can be considerable even at low grafting densities, making
it easy the system to access the brush regime. The use of charged
polymers instead of uncharged polymers introduces additional
length scales such as Bjerrum length and Debye screening length to
the system. Polyelectrolyte brushes also have been subjected to
extensive investigations involving both theoretical
\cite{Pincus,Zhulina1,Zhulina2,Borisov2,Zhulina3,Naji1,Ahrens,Naji2}
and computer simulation
 methods \cite{Naji1,Ahrens,Naji2,Csajka,Seidel,Fazli}. At high enough grafting densities
and charge fractions of polyelectrolyte chains, most of
counterions are trapped inside the polyelectrolyte brush and
competition between osmotic pressure of the counterions and
elasticity of the chains determines the brush thickness. This
regime of a polyelectrolyte brush is known as the osmotic regime
in which some theoretical scaling methods predict no dependence of
the brush thickness to the grafting density
\cite{Pincus,Borisov1}. However, other  scaling method that takes
into account the excluded volume effects and nonlinear elasticity
of polyelectrolyte chains predicts a linear dependence of the
brush thickness on the grafting density and is in agreement with
experiment and simulation \cite{Naji1,Ahrens,Naji2}. Also, it has
been shown that diffusion of a fraction of counterions outside the
polyelectrolyte brush leads to a logarithmic dependence of the
average brush thickness on the grafting density \cite{Zhulina3}.
\begin{figure}[t]
\includegraphics[width=0.95 \columnwidth]{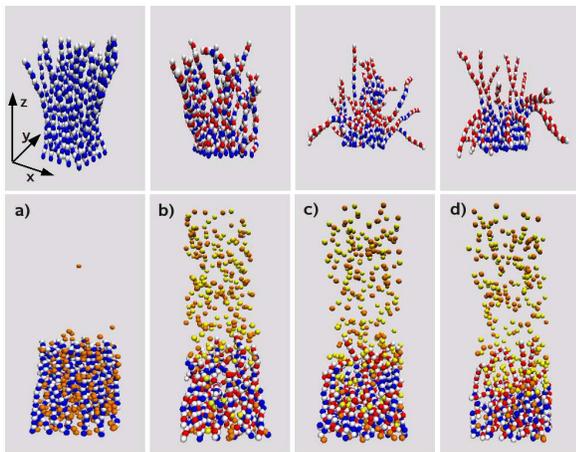}
\caption{(Color online) Typical equilibrium configurations of
brushes formed by a)PE, b)PA11, c)PA33, and d)PA66 chains at
grafting density $\rho_a\sigma^2=0.1$. In the upper plans,
counterions are not shown and periodic boundary condition is
removed for clarity. As it can be seen, in the brushes of PA33 and
PA66, the chains are buckled toward the grafting surface. }
\label{snapshots}
\end{figure}

Brushes formed by grafted diblock polyampholytes have also been
investigated by lattice mean field modeling
\cite{Shusharina2,Shusharina3} and computer simulation
\cite{Akinchina1}. The effect of chain stiffness, charge density,
and grafting density on spherical brushes of diblock
polyampholytes and interaction between colloids with grafted
diblock polyampholytes have been studied using Monte Carlo
simulations \cite{Akinchina2,Linse}. Despite the case of an
osmotic polyelectrolyte brush that counterions are trapped inside
the brush and have a crucial rule in the equilibrium brush
thickness, in the case of a polyampholyte brush most of
counterions are outside and equilibrium brush thickness is mainly
determined by the chains properties.

A brush of semiflexible charged polymers is a dense assembly of
such polymers in which the interplay between electrostatic
correlations, entropic effects and the bending elasticity of the
chains can bring about a variety of equilibrium properties. The
strength of the electrostatic correlations in a polyampholyte
brush depends on the sequence of charged monomers along the
chains. With similarly charged monomers being arranged in long
blocks, the electrostatic correlations are of great importance in
the system. It seems that for a brush formed by semiflexible
polyampholyte chains, each chain containing equal number of
positive and negative monomers, studying the effect of monomers
sequence on equilibrium properties of the brush could be an step
toward better understanding of such system s.

In this paper, we study planar brushes of semiflexible
polyampholytes using extensive molecular dynamics (MD) simulations
in a wide range of the grafting density and at various salt
concentrations. In our simulations, each end-grafted polyampholyte
chain contains equal number of positively and negatively charged
monomers (isoelectric condition). We investigate the effect of the
sequence of charged monomers along the chains on equilibrium
properties of the brush. Charged monomers of each sign are
arranged in blocks of equal length and monomers of opposite sign
form regular alternation of positive and negative blocks (see Fig.
\ref{structures} b, c and d). Different charge sequences in our
simulations correspond to different lengths of above mentioned
blocks. We also consider the case of polyelectrolyte brush in our
studies in which all the charged monomers of the chains are of the
same sign for comparison.

We find that at each value of the grafting density, the brush
formed by polyampholytes with longer blocks of similarly charged
monomers has smaller average thickness, i.e., the lowest value of
the average thickness corresponds to the brush of diblock
polyampholytes. The values of the grafting density that we use in
our simulations correspond to the osmotic regime of the
polyelectrolyte brush. In this regime, most of the counterions are
contained inside the polyelectrolyte brush and the average brush
thickness has a weak linear dependence on the grafting density.
Dependence of the average thickness of polyelectrolyte brush that
we obtain from our simulations is in agreement with nonlinear
theory of ref. \cite{Naji1}. In the case of the polyampholyte
brush, we find that the average thickness considerably varies as a
function of the grafting density and with longer blocks of
similarly charged monomers, this dependence is stronger.
Counterions of a polyampholyte brush are not trapped inside the
brush and dependence of the average thickness on the grafting
density originates from a different mechanism than that of a
polyelectrolyte brush (see the text). We also study the
equilibrium statistics of conformations of the chains at various
grafting densities and salt concentrations both for polyampholyte
and polyelectrolyte brushes. We find that in the case of
polyampholyte brushes with long blocks of similarly charged
monomers,  at low grafting densities the inter-chain electrostatic
interactions link the chains to each other in the vicinity of the
grafting surface and cause the brush to collapse. At high grafting
densities, although the excluded volume interactions tend to
increase the brush thickness, the electrostatic interaction
dominate over the bending elasticity of the chains and lead them
to buckle and the brush thickness to decrease. Also it is observed
that with long blocks of similarly charged monomers along the
chains, the average thickness of the polyampholyte brush is an
increasing function of the salt concentration.

The rest of the paper is organized as follows. In Sec. \ref{sim}
we describe our model and details of molecular dynamics
simulations. We present results of the simulations in detail and
concluding remarks in Sec. \ref{result}. Finally, in Sec.
\ref{summary} we summarize the paper.

\section{The Model and the Simulation Method}    \label{sim}

In our simulations which are performed with the MD simulation
package ESPResSo \cite{ESPResSo}, each brush is modeled by $M$
semiflexible polyelectrolyte/polyampholyte bead-spring chains of
length $N$ ($N$ spherical monomers) which are end-grafted onto an
uncharged surface at $z=0$.
\begin{figure}[t]
\includegraphics[width=0.90 \columnwidth]{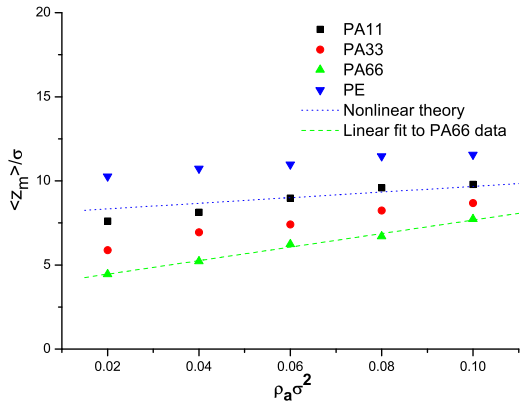}
\caption{(Color online) Average thickness of brushes formed by PE,
PA11, PA33 and PA66 chains versus dimensionless grafting density
$\rho_a\sigma^2$ with no added salt. Dotted line is the prediction
of the scaling theory of ref. \cite{Naji1} with
$\sigma_{eff}\simeq 1.02\sigma$ and dashed line is a linear fit to
our data for the brush of PA66 chains. As it can be seen
dependence of the average thickness of polyampholyte brushes with
longer blocks of similarly charged monomers on the grafting
density is stronger. The size of the symbols corresponds to the
size of error bars.}\label{thickness}
\end{figure}
Short-range repulsive interaction which is described by a shifted
Lennard-Jones potential,
\begin{equation}
 u_{LJ}(r) = \left\lbrace
  \begin{array}{l l}
    4\varepsilon\
    \{(\frac{\sigma}{r})^{12}-(\frac{\sigma}{r})^{6}+\frac{1}{4}\} & \text{if $r<r_{c}$},\\
    0 & \text{if $r \geq r_{c}$},
  \end{array}
\right. \label{slj}
\end{equation}
is considered between monomers in which $\epsilon$ and $\sigma$
are the usual Lennard-Jones parameters and the cutoff radius is
$r_{c}=2^{1/6} \sigma$. Neighboring beads along each chain are
bonded to each other by a FENE (finite extensible nonlinear
elastic) potential \cite{Grest},
\begin{equation}
u_{bond}(r)=\left\lbrace
     \begin{array}{l l}
       - \frac{1}{2}k_{bond}R_{0}^{2}\ln(1-(\frac{r}{R_{0}})^{2})& \text{if $r<R_{0}$},\\
       0 & \text{if $r \geq R_{0}$},
  \end{array}
\right. \label{FENE}
\end{equation}
with bond strength $k_{bond}=30\varepsilon/\sigma^{2}$ and maximum
bond length $R_{0}= 1.5\sigma$. The chains are semiflexible and
their elasticity is modeled by a bond angle potential,
\begin{equation}
u_{bend}(r)=k_{bend}(1-\cos\theta) \label{angle},
\end{equation}
in which $\theta$ is the angle between two successive bond vectors
of a chain. In our simulations $k_{bend} = 25k_BT$ which means
that $l_p=25\sigma\simeq L_c$ in which $l_p$ and $L_c$ are the
persistence length and the contour length of the chains
respectively. The simulation box is of volume $L \times L \times
L_z$ in which $L$ is the box width in $x$ and $y$ directions and
$L_z$ is its height in $z$ direction and the grafting density is
given by $\rho_{a}=M/L^{2}$. The positions of anchored monomers
which are fixed during the simulation, form an square lattice with
lattice spacing $d=\rho_{a}^{-1/2}$ on the grafting surface ($x-y$
plane). All particles interact repulsively with the grafting
surface at short distances with the shifted Lennard-Jones
potential introduced in Eq. \ref{slj}. In addition, a similar
repulsive potential is applied at the top boundary of the
simulation box and in our simulations $L_z =2N\sigma$. We model
positive and negative ions of monovalent salt by equal number of
spherical Lennard-Jones particles of diameter $\sigma$ with
charges $e$ and $-e$ respectively. All the charged particles
interact with the Coulomb interaction

\begin{equation}
u_{C}(r)= k_{B}Tq_{i}q_{j}\frac{l_{B}}{r} \label{coul}
\end{equation}
\begin{figure}[t]
\includegraphics[width=0.80\columnwidth]{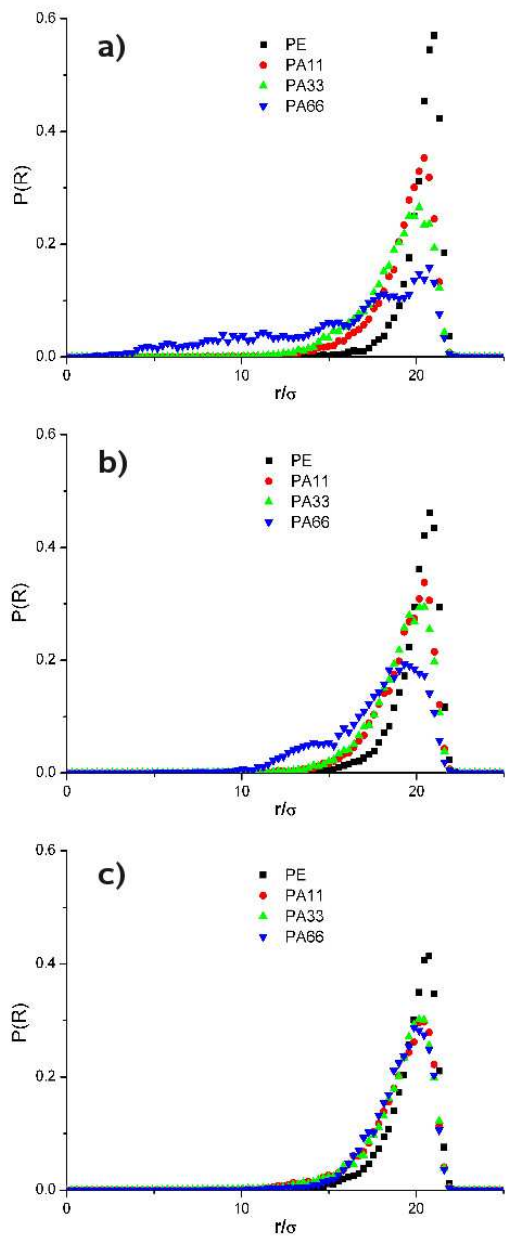}
\caption{ The histogram of the average end-to-end distance of the
chains for brushes formed by PE, PA11, PA33 and PA66 chains at
grafting densities a) $\rho_{a}\sigma^{2}=0.1$,
b)$\rho_{a}\sigma^{2}=0.06$ and c) $\rho_{a}\sigma^{2}=0.02$. }
\label{p_r}
\end{figure}
in which $q_{i}$ and $q_{j}$ are charges of particles $i$ and $j$
in units of elementary charge $e$ and $r$ is separation between
them. The Bjerrum length, $l_B$, which determines the strength of
the Coulomb interaction relative to the thermal energy, $k_BT$, is
given by $l_{B}=e^{2}/\varepsilon k_{B}T$, where
 $\varepsilon$ is the dielectric constant of the solvent. $l_{B}=2\sigma$ in our
simulations. We apply periodic boundary conditions only in two
dimensions ($x$ and $y$). To calculate Coulomb forces and
energies,
 we use the so-called $MMM$
technique introduced by Strebel and Sperb \cite{Strebel} and
modified for laterally periodic systems ($MMM2D$) by Arnold and
Holm \cite{Arnold}. The temperature in our simulations is kept
fixed at $k_{B}T=1.2\varepsilon$ using a Langevin thermostat.

In our simulations we consider brushes containing $M=25$ anchored
chains, each chain consisting of $N=24$ monomers, $fN$ of them are
charged and $f=\frac{1}{2}$. We also consider $\frac{ M \times
N}{2}$ counterions to neutralize the chains charge. The following
sequences of charged monomers along the chains are considered to
model polyelectrolyte and three different polyampholyte brushes
respectively: PE (\ldots - - - - \ldots) PA11 (\ldots + - + - + -
\ldots), PA33 (\ldots + + + - - - \ldots)and PA66 (\ldots + + + +
+ + - - - - - - \ldots). Each chain consist of an alternating
sequence of charged and neutral monomers and the neutral monomers
are not shown for simplicity (see Fig. \ref{structures}). We do
simulations of the brushes formed by four above mentioned
different chains at dimensionless grafting densities
$\rho_{a}\sigma^{2}=0.02,0.04,0.06,0.08,0.10$ and at various
monovalent salt concentrations changing from $c_s\sigma^3=0$ to
$c_s\sigma^3=0.22$. In the beginning of each simulation, all of
the chains are straight and perpendicular to the grafting surface
and all the ions are randomly distributed inside the simulation
box. We equilibrate the system for $1.6\times 10^6$ MD time steps
which is enough for all values of the grafting density mentioned
above and then calculate thermal averages over $1500$ independent
configurations of the system selected from $2.25\times 10^6$
additional MD steps after equilibration. MD time step in our
simulations is $\tau=0.01\tau_0$ in which
$\tau_0=\sqrt{\frac{m\sigma^2}{\varepsilon}}$ is the MD time scale
and $m$ is the mass of the particles.

We calculate the average brush thickness which can be measured by
taking the first moment of the monomer density profile
\begin{equation}
\langle z_{m} \rangle =\frac{\int_{0}^{\infty}{z
\rho_{m}(z)}{dz}}{\int_{0}^{\infty}{ \rho_{m}(z)}{dz}},\label{z_m}
\end{equation}
in which $\rho_{m}(z)$ is the number density of monomers as a
function of the distance from the grafting surface. For better
understanding of the statistics of the chains conformations, we
calculate the histogram of the mean end-to-end distance of the
chains, $P(R)$, in which $R=\frac{1}{M}\sum_{i=1}^{M}|\vec{R_i}|$
and $\vec{R_i}$ is the end-to-end vector of chain $i$. We also
calculate the histogram of the average distance of the end
monomers of the chains from the grafting surface, $P(z_{end})$, in
which $z_{end}=\frac{1}{M}\sum_{i=1}^{M}z_i$ and $z_i$ is the $z$
component of the end monomer of chain $i$.
\begin{figure}[t]
\includegraphics[width=0.80\columnwidth]{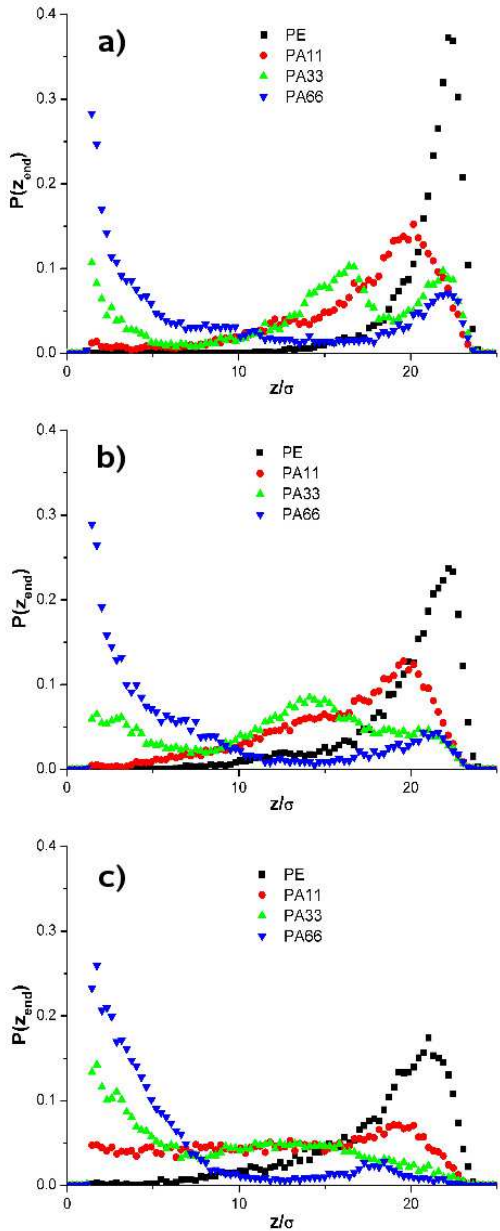}
\caption{ The histogram of the average distance of the end
monomers of the chains from the grafting surface for brushes
formed by PE, PA11, PA33 and PA66 chains at grafting densities a)
$\rho_{a}\sigma^{2}=0.1$, b)$\rho_{a}\sigma^{2}=0.06$ and c)
$\rho_{a}\sigma^{2}=0.02$.} \label{p_z}
\end{figure}
Also, as a measure of the lateral fluctuations of the chains we
define $l_{lat}$ as $l_{lat}=\frac{1}{M}\sum_{i=1}^{M}R_{i||}$ in
which $R_{i||}=|\vec{R_i}-\vec{R_i}.\vec{\hat{z}}|$ is the
magnitude of the lateral component of $\vec{R_i}$.

\section{Results and Concluding Remarks} \label{result}

We calculate the average thickness of the brushes formed by
polyelectrolyte chain, PE, and three different polyampholyte
chains PA11, PA33 and PA66 at different grafting densities. In
Fig. \ref{thickness} the average brush thickness versus
dimensionless grafting density $\rho_a\sigma^2$ with no added salt
is shown.  As it can be seen in this figure, the average thickness
of the polyelectrolyte brush has a week dependence on the grafting
density. With grafting densities that we use in our simulations,
the Gouy-Chapman length of the polyelectrolyte brush which is
defined as $\lambda_G=(2\pi \rho_afNl_B)^{-1}$, is quite smaller
than the length of the chains $(\lambda_G < \sigma)$ and most of
the counterions are contained inside the brush (see Fig.
\ref{structures} a). In this regime of polyelectrolyte brushes,
the osmotic pressure of the counterions determines the thickness
of the brush and the week dependence of the brush thickness on the
grafting density is expected. In fig. \ref{thickness} the
prediction of the scaling theory of ref. \cite{Naji1} (see Eq.
(28) of \cite{Naji1}) corresponding to our parameters is shown for
comparison. As it can be seen, this scaling theory describes well
the dependence of the average thickness of semiflexible osmotic
polyelectrolyte brush on the grafting density obtained from our
simulations.

In the case of the brushes of polyampholyte chains, the net charge
of the brush layer is zero and most of the counterions are outside
the brush. The average thickness in this case has a considerable
dependence on the grafting density and the interaction between the
anchored chains. At each value of the grafting density, brush
thickness decreases with increasing the length of the blocks of
similarly charged monomers. Also, as it can be seen in Fig.
\ref{thickness}, the average thickness of the brushes formed by
the chains with longer blocks of similarly charged monomers has
stronger dependence on the grafting density (see growing slope of
$\frac{\langle z_m \rangle }{\sigma}$versus $\rho_a\sigma^2$ from
PA11 to PA66).

To describe such dependence of the brushes thickness on the
grafting density it is instructive to study the equilibrium
statistics of conformations of the chains. Accordingly, we monitor
the histograms $P(R)$ and $P(z_{end})$ which are defined in Sec.
\ref{sim} in our simulations. In Figs. \ref{p_r} and \ref{p_z} the
histograms $P(R)$ and $P(z_{end})$ at three different grafting
densities are shown for brushes of polyelectrolyte and three
different polyampholyte chains. It can be understood from Figs.
\ref{p_r} and \ref{p_z} that decreasing the average thickness of a
polyampholyte brush with increasing the length of the blocks of
similarly charged monomers originates from two different
mechanisms at low and high grafting densities. At low grafting
densities, the anchored chains both in polyelectrolyte brush and
in three different polyampholyte brushes have stretched
conformations and the average end-to-end distance, $R$, has its
maximum value in most of equilibrium configurations (see Fig.
\ref{p_r} c). In Fig. \ref{p_z} c, the histogram $P(z_{end})$
shows that at such low values of the grafting density the
polyelectrolyte brush is aligned and the PE chains are mostly
perpendicular to the grafting surface. In this figure, it can be
seen that in the case of polyampholyte brushes with long blocks of
similarly charged monomers, $z=0$ part of $P(z_{end})$ is nonzero
and increases with increasing the length of blocks of similarly
charged monomers. From these two histograms one can conclude that
in the brushes of polyampholytes with long blocks of similarly
charged monomers, PA33 and PA66, the chains have rodlike
conformations and fluctuate in the vicinity of the grafting
surface. Snapshots of the brushes of these chains (not shown here)
 show that the chains are linked to each other as
doublets near the grafting surface because of the strong
electrostatic attractions. In the case of the brush of PA11
chains, the end monomers of the chains have a uniform distribution
inside the brush. In this case, electrostatic neutrality is
satisfied in smaller length scales and the electrostatic
correlations between the chains are very weak.
\begin{figure}[t]
\includegraphics[width=0.90\columnwidth]{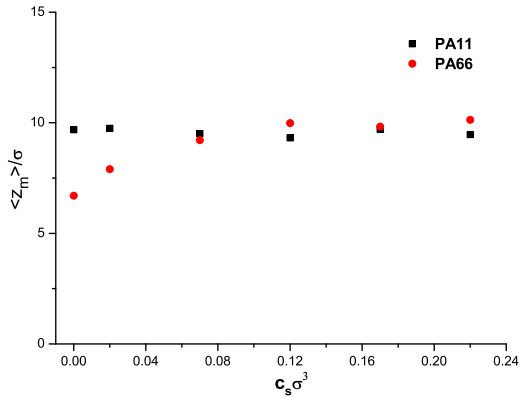}
\caption{ Salt concentration dependence of the average thickness
of brushes of PA11 and PA66 chains at grafting density
$\rho_{a}\sigma^{2}=0.08$. As it can be seen, the thickness of the
brush formed by the chains with long blocks of similarly charged
monomers is an increasing function of the salt concentration. The
size of the symbols corresponds to the size of error bars.}
\label{thickness_salt}
\end{figure}
\begin{figure}[t]
\includegraphics[width=0.75\columnwidth]{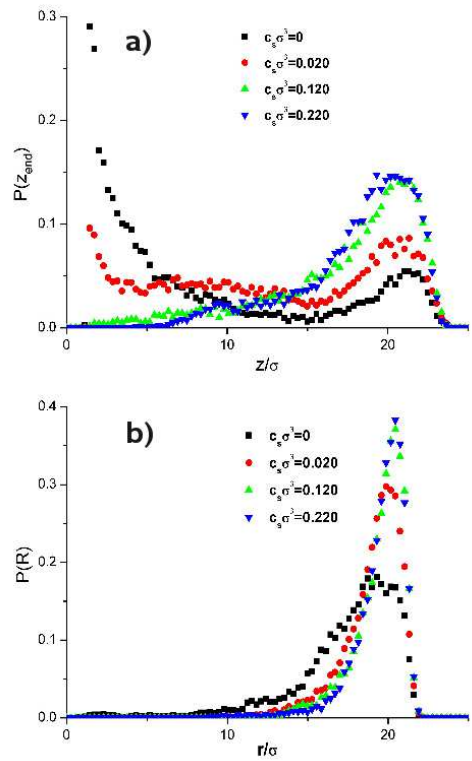}
\caption{Histograms a)$P(z_{end})$ and b)$P(R)$ for the brush
formed by PA66 chains at different salt concentrations. The
grafting density is $\rho_{a}\sigma^{2}=0.08$.  }
\label{salt_p_r_z}
\end{figure}

At high grafting densities, the histogram of the average
end-to-end distance
 of the chains shows
that in the brushes of polyampholyte chains consisting of long
blocks of similarly charged monomers, despite the case of low
grafting density, the chains tend to have buckled conformations.
The histogram of the average $z$ component of the end monomers,
$P(z_{end})$, shows two maximums around $z=0$ and $z=L$ in the
case of the brush of diblock PA66 chains and three maximums in the
case of the brush of PA33 chains. The energy needed for buckling a
semiflexible chain $(L_c\sim l_p)$, $E_b$, is of the order of
thermal energy, $k_BT$. In dense brushes of the chains PA33 and
PA66, blocks of similarly charged monomers of the chains form
highly charged layers of monomers which produce strong
electrostatic field. When the electrostatic energy of similarly
charged monomers of the end blocks of the chains due to the
electric field of the layer of opposite charge exceeds the bending
energy, $E_b\sim k_BT$, the chains buckle toward the grafting
surface. For example, in the case of the brush of diblock PA66
chains, the electrostatic energy of positively charged monomers of
each chain roughly can be written as $E_{el}\simeq
\frac{fNe}{2}E\frac{L}{2}$ in which
$E=4\pi\rho_ae/\varepsilon=4\pi k_BTl_B\rho_a/e$ is approximately
the electric field of the layer of anchored blocks of negatively
charged monomers. With parameters that we use in our simulations,
at high grafting densities of the brushes formed by PA33 and PA66
chains, $E_{el}>k_BT$ and buckling of the chains is expected. In
the case of the brush of PA66 chains, two maximums in the
histogram $P(z_{end})$ shows that buckled chains and those that
are straight and perpendicular to the grafting surface coexist. In
the brush of PA33 chains, there are two layers of negatively
charged monomers and considering the chains of straight
configuration, three maximums seen in the histogram $P(z_{end})$
is expected.

The effect of added salt on the brushes of polyampholyte chains is
also studied using explicit monovalent salt ions. We observe that
properties of the brushes formed by the chains containing longer
blocks of similarly charged monomers are more sensitive to the
salt concentration. The average thickness of these brushes at all
grafting densities increase with increasing the salt concentration
and reach to its maximum value (see Fig. \ref{thickness_salt}).
This behavior is completely different from that of a
polyelectrolyte brush which its thickness is known that decreases
with increasing the salt concentration. In the case of a
polyelectrolyte brush the electrostatic repulsion between the
chains tend to swell the brush and the screening effect of added
salt decreases the average brush thickness. However, in the case
of the brush of polyampholyte chains with long blocks of similarly
charged monomers, the electrostatic interactions between the
chains are attractive and tend to decrease the brush thickness. In
these brushes,  electrostatic screening of the salt ions increases
the average thickness. It also can be seen in Fig.
\ref{thickness_salt} that the added salt doesn't change the
average thickness of the brush of PA11 chains in which because of
local electrostatic neutrality the electrostatic correlations are
negligible. In Fig. \ref{salt_p_r_z} the equilibrium histograms
$P(R)$ and $P(z_{end})$ for the brush of PA66 chains are shown at
different salt concentrations. It can be seen that with adding
more salt to the system, the chains become more straight (see part
b of the figure) and $z=0$ part of the histogram $P(z_{end})$
vanishes which means that the brush becomes more aligned.

To estimate the importance of finite size effect in our
simulations, we calculated equilibrium average of $l_{lat}$
(defined in Sec. \ref{sim}) for the brush of PA66 chains at
highest grafting density $\rho_a\sigma^2=0.1$ and found that
$\langle l_{lat}\rangle\simeq11.7\sigma$. The fact that the value
of $\langle l_{lat} \rangle$ is less than the lateral box length,
$L\simeq 16\sigma$, means that the chains doesn't overlap with
their own images in our simulations. We also repeated the
simulation of the PA66 brush at grafting density
$\rho_a\sigma^2=0.1$ with $M=64$ chains (instead of $M=25$) and
didn't obtain different results.

\section{Discussion} \label{discussion}

Osmotic pressure of trapped counterions inside polyelectrolyte
brush in the osmotic regime is one of the main factors determining
the equilibrium brush thickness. Nonetheless, because of the
electrostatic neutrality of the chains, Gouy-Chapman length
diverges in the case of polyampholyte brushes considered in this
paper and most of counterions are outside the brush layer (see
Fig. \ref{snapshots} b, c and d). These counterions have no rule
in the equilibrium brush thickness. In this case the brush
thickness is resulted from the interplay between inter- and
intra-chain electrostatic attraction of oppositely charged blocks,
inter-chain excluded volume interactions and the bending
elasticity of the chains. Hence, dependence of the average
thickness of such polyampholyte brushes on the grafting density is
of quite different mechanism. However, it is interesting that this
dependence is still linear similar to that of a polyelectrolyte
brush (see Fig. \ref{thickness}). This dependence in brushes of
polyampholyte chains with long blocks of similarly charged
monomers is appreciably stronger than the case of polyelectrolyte
brush (for example in Fig. \ref{thickness} the slopes of dotted
and dashed lines are 16.7 and 40.2 respectively). An important
parameter in a brush of semiflexible polyampholytes is the bending
rigidity of the chains, $k_{bend}$. With decreasing the value of
$k_{bend}$, electrostatic attractions dominate over the chains
stiffness and in competition with excluded volume effects
determine the average brush thickness. The opposite extreme is the
case of the brush of rod-like polyampholytes in which
electrostatic correlations can not buckle the chains. So,
different regimes corresponding to different values of $k_{bend}$
should be investigated \cite{Baratlo}.

\section{Summary} \label{summary}

In summary, we have used molecular dynamics simulations to study
planar brushes formed by semiflexible polyelectrolytes and
polyampholyte chains with different sequences of charged monomers
at various grafting densities and salt concentrations. It has been
shown that at grafting densities corresponding to the osmotic
regime of the polyelectrolyte brush, the average brush thickness
is a weak linear function of the grafting density in agreement
with predictions of ref. \cite{Naji1}. The average thickness and
equilibrium properties of the polyampholyte brushes have
considerable dependence on the grafting density and the salt
concentration. This dependence is stronger for brushes of
polyampholyte chains containing longer blocks of similarly charged
monomers. In brushes of polyampholyte chains with long blocks of
similarly charged monomers, at low grafting densities the
electrostatic attractions link the chains to each other in the
vicinity of the grafting surface and collapse the brush. At high
grafting densities, the electrostatic correlations dominate over
the bending rigidity of the chains and cause them to buckle toward
the grafting surface. Despite the case of polyelectrolyte brushes,
the average thickness of polyampholyte brushes with long blocks of
similarly charged monomers increases with increasing the salt
concentration.

We are grateful to A. Naji, R. Golestanian and F. Mohammad-Rafiee
for useful comments and discussions.


\begin{thebibliography}{99}

\bibitem{Hong}
Y. Hong, R. L. Legge, S. Zhang, P. Chen, Biomacromolecules.
\textbf{4},  1434 (2003).

\bibitem{Jun}
S. Jun, Y. Hong, H. Imamura, B.-Y. Ha, J. Bechhoefer, P. Chen
Biophys. J. \textbf{87},  1249 (2004).

\bibitem{Messina}
R. Messina Eur. Phys. J. E \textbf{22} , 325 (2007).

\bibitem{rev_Dobrynin}
A. V. Dobrynin, R. H. Colby, M. Rubinstein, J. Polym. Sci., Part
B: Polym. Phys. \textbf{42},  3513 (2004).

\bibitem{Imbert}
J. B. Imbert, J. M. Victor, N. Tsunekawa, Y. Hiwatari, Phys. Lett.
A \textbf{258},  92 (1999).

\bibitem{Baumketner}
A. Baumketner, H. Shimizu, M. Isobe, Y. Hiwatari, J. Phys.:
Condens. Matter \textbf{13},  10279 (2001).

\bibitem{Wang}
Z. Wang, M. Rubinstein, Macromolecules \textbf{39},  5897 (2006).

\bibitem{Castelnovo}
M. Castelnovo, J. F. Joanny, Macromolecules \textbf{35}, 4531
(2002).

\bibitem{Shusharina1}
N. P. Shusharina, E. B. Zhulina, A. V. Dobrynin, M. Rubinstein,
Macromolecules \textbf{38},  8870 (2005).

\bibitem{Zheng}
G. Z. Zheng, G. Meshitsuka, A. Ishizu,  J Polym Sci Part B, Polym
Phys \textbf{33},  867 (1995).

\bibitem{Ehrlich}
G. Ehrlich, P. Doty, J Am Chem Soc \textbf{76},  3764 (1954).

\bibitem{McCormick}
C. L. McCormick, L. C. Salazar, Macromolecules \textbf{25}, 1896
(1992).

\bibitem{Corpart}
J. M. Corpart, F. Candau, Macromolecules \textbf{26},  1333
(1993).

\bibitem{Wijmans}
C. M. Wijmans, E. B. Zhulina, Macromolecules \textbf{26}, 7214
(1993).

\bibitem{Lindberg}
E. Lindberg, C. J. Elvingson, Chem. Phys. \textbf{114}, 6343
(2001).

\bibitem{Klos}
J. Klos, T. Pakula, Macromolecules \textbf{37},  8145 (2004).

\bibitem{Almusallam}
A. S. Almusallam, D. S. Sholl, Nanotechnology \textbf{16}, S409
(2005).

\bibitem{Pincus}
P. Pincus, Macromolecules \textbf{24},  2912 (1991).

\bibitem{Zhulina1}
E. B. Zhulina, O. V. Borisov, Macromolecules \textbf{29}, 2618
(1996).

\bibitem{Zhulina2}
E. B. Zhulina, J. K. Wolterink, O. V. Borisov, Macromolecules
\textbf{33},  4945 (2000).

\bibitem{Borisov2}
O.V. Borisov, T.M. Birstein, E.B. Zhulina, J. Phys. II \textbf{2},
63
 (1992).

\bibitem{Zhulina3}
E.B. Zhulina, O.V. Borisov, J. Chem. Phys. \textbf{107}, 5952
(1997).

\bibitem{Naji1}
A. Naji, R.R. Netz, C. Seidel Eur. Phys. J. E. \textbf{12}, 223
(2003).

\bibitem{Ahrens}
H. Ahrens, S. Forster, C.A. Helm, N.A. Kumar, A. Naji, R.R. Netz,
C. Seidel J. Phys. Chem. B \textbf{108}, 16870 (2004).

\bibitem{Naji2}
A. Naji, C. Seidel, R.R. Netz Adv. Polym. Sci.  \textbf{198}, 149
(2006).

\bibitem{Csajka}
F. S. Csajka, C. Seidel, Macromolecules \textbf{33},  2728 (2000).

\bibitem{Seidel}
C. Seidel, Macromolecules \textbf{36},  2536 (2003).


\bibitem{Fazli}
H. Fazli, R. Golestanian, P. L. Hansen, M. R. Kolahchi, Eur. Phys.
Lett \textbf{73},  429 (2006).


\bibitem{Borisov1}
O.V. Borisov, T.M. Birstein, E.B. Zhulina, J. Phys. II \textbf{1},
521 (1991).



\bibitem{Shusharina2}
N. P. Shusharina, P. Linse, Eur. Phys. J. E \textbf{4}, 399
(2001).

\bibitem{Shusharina3}
N. P. Shusharina, P. Linse, Eur. Phys. J. E \textbf{6}, 147
(2001).

\bibitem{Akinchina1}
A. Akinchina, N. P. Shusharina, P. Linse, Langmuir \textbf{20},
 10351 (2004).

\bibitem{Akinchina2}
A. Akinchina, P. Linse, Langmuir \textbf{23}, 1465 (2007).

\bibitem{Linse}
P. Linse, J. Chem. Phys \textbf{126}, 114903 (2007).

\bibitem{ESPResSo}
H.J. Limbach, A. Arnold, B.A. Mann and C. Holm, Comp. Phys.
Communications \textbf{174}, 704 (2006).

\bibitem{Grest}
G.S. Grest and K. Kremer, Phys. Rev. A \textbf{33}, 3628  (1986).

\bibitem{Strebel}
R. Strebel, R. Sperb, Mol. Simul. \textbf{27}, 61 (2001).

\bibitem{Arnold}
A. Arnold, C. Holm, Comput. Phys. Commun. \textbf{148}, 327
(2002).

\bibitem{Baratlo}
M. Baratlo, H. Fazli (unpublished).


\end{thebibliography}
\end{document}